\documentstyle[11pt]{article}

\makeatletter
\@addtoreset{equation}{section}
\makeatother

\def\dalemb#1#2{{\vbox{\hrule height .#2pt
        \hbox{\vrule width.#2pt height#1pt \kern#1pt
                \vrule width.#2pt}
        \hrule height.#2pt}}}

\def\gtlt{\mathrel{\raise4.5pt\hbox{\oalign{$\scriptstyle>$\crcr
$\scriptstyle<$}}}}

\def\ie{{\it i.e.\ }}

\newcommand{\be}{\begin{equation}}
\newcommand{\ee}{\end{equation}}
\newcommand{\bea}{\begin{eqnarray}}
\newcommand{\eea}{\end{eqnarray}}

\def\ft#1#2{{\textstyle{{\scriptstyle #1}\over {\scriptstyle #2}}}}
\def\fft#1#2{{#1 \over #2}}

\def\half{{\textstyle{1\over2}}}

\newcommand{\hoch}[1]{$\, ^{#1}$}

\newcommand{\auth}{
M.J. Duff\hoch{\dagger1},
James T. Liu\hoch{\dagger},
K.S. Stelle\hoch{\ddagger2}}

\begin{document}
\begin{flushright}
\hfill{UM-TH-00-15} \\
\hfill{Imperial/TP/99-0/33} \\
\hfill{\bf hep-th/0007120}\\
\end{flushright}

\vspace{10pt}

\begin{center}

{\large {\bf A supersymmetric Type IIB Randall-Sundrum realization}}

\vspace{20pt}

\auth

\vspace{10pt}
{\hoch{\dagger} \it
Randall Laboratory, Department of Physics, University of Michigan,\\
Ann Arbor, MI 48109--1120}

\vspace{10pt}
{\hoch{\ddagger} \it
The Blackett Laboratory, Imperial College\\
Prince Consort Road, London SW7 2BZ, UK}

\vspace{30pt}

\underline{ABSTRACT}
\end{center}

We show that an earlier domain wall solution of type IIB 
supergravity provides a supersymmetric realization of the 
Randall-Sundrum brane-world, and give its ten-dimensional 
interpretation in terms of IIB 3-branes. We also explain how 
previous no-go theorems are circumvented. In particular, whereas 
$D=5$ supergravity scalars have AdS$_{5}$ energy $E_{0}\leq 4$ and are 
unable to support a $D=5$ positive tension brane, our scalar has 
$E_{0}=8$, and is the breathing mode of the $S^{5}$ compactification.
Another essential element of the construction is the implementation of 
a $Z_{2}$ symmetry by patching together compactifications with 
opposite signs for their 5-form field strengths. This is thus a IIB 
analogue of a previous $D=5$ 3-brane realization of the Ho\v{r}ava-Witten 
orbifold. A mode-locking phenomenon avoids the appearance of negative 
energy zero-modes in spite of the necessity of a $D=10$ negative 
tension brane-source.

{\vfill\leftline{}\vfill
\vskip 10pt \footnoterule
{\footnotesize \hoch{1}
Research supported in part by NSF Grant PHY-9722090
\vskip  -12pt} \vskip   14pt

{\footnotesize \hoch{2}
Research supported in part by PPARC under SPG grant 613.
}
}

\pagebreak
\setcounter{page}{1}

\section{Introduction}

The purpose of this paper is first to show that the type IIB  
domain wall solution of Bremer {\it et al.}~\cite{Bremer} provides a 
supersymmetric realization of the Randall-Sundrum brane-world 
\cite{RS1,RS2} and secondly to give its ten-dimensional interpretation 
in terms of IIB 3-branes.   

The idea that our universe may be a 3-brane in a higher dimensional
spacetime has a history going back nearly two decades
\cite{Akama,Rubakov:1983bb,Rubakov:1983bz,Gibbons:1988ps,%
Horowitz:1991cd,Duff:1991pe}.
More recently, another viewpoint on this basic idea has grown out
of the Ho\v{r}ava-Witten \cite{Horava:1996qa,Horava:1996ma} model
for M--theory/heterotic string duality, based upon an $S^1/Z_2$
orbifold in $D=11$ spacetime. This orbifold construction was later
realized in a $D=5$ compactification by a concrete solution to semiclassical
M--theory, \ie $D=11$ supergravity \cite{losw1,losw2}. A key point in this
construction was the introduction of flux for the M--theory 4-form field strength
$G_{[4]}$ wound around the compact dimensions, which were taken to be a
Calabi-Yau 3-fold. The resulting reduced theory is a specific version of
matter-coupled $D=5$, $N=2$ supergravity. This dimensionally-reduced
theory has a scalar potential arising from the $G_{[4]}$ flux, which rules out
flat space or indeed any maximally symmetric space as a solution to the equations
of motion.  But this $D=5$ reduced theory readily admits
domain wall, \ie 3-brane, solutions. A natural configuration is a pair of
two 3-branes in a $Z_2$ symmetric configuration; projecting the fields
of this theory into the subspace of $Z_2$ invariant configurations then
reproduces the Ho\v{r}ava-Witten orbifold. As in the original $D=11$/$D=10$
discussion, the massless brane-wave excitations of this scenario are
not easily deduced by direct analysis of the solution, but one may
obtain information about the zero-modes by anomaly inflow arguments.
These may either be carried out in $D=10$, leading to the original
Ho\v{r}ava-Witten prediction of a $D=10$, $N=1$ super Yang-Mills $E_8$ gauge
multiplet residing on each of the two fixed planes of the orbifold,
with the resulting structure subsequently reduced to $D=5$, or one may
carry out the anomaly analysis directly in $D=5$, yielding more general
possibilities for gauge structure \cite{ls}.

Another theory in which similar constructions can be made is $D=10$
type IIB supergravity. This has a self-dual 5-form field strength that
supports the D3-brane, which is the basis for much recent discussion of
the Maldacena conjecture, linking string theory in the near-horizon region
of the D3-brane to a Yang-Mills theory quantized on the boundary of the
associated asymptotic anti de Sitter space, which is the near-horizon limiting
spacetime. In the pure supergravity context, relations between $p$-branes
in higher dimensions and domain walls arising after dimensional reduction
on spheres was developed in \cite{Bremer}, including the case of the
D3-brane of type IIB theory.

Meanwhile, another development was brewing. Randall and Sundrum \cite{RS1}
proposed a simple model of physics on 3-branes embedded in $D=5$ anti de
Sitter space, first in a model with two 3-branes, one of positive and one
of negative tension. This model was criticized for the apparent danger
of non-physical modes from the negative tension brane, and also because
the modulus related to the distance between the two branes gave another
parameter needing fixing in any phenomenological analysis. Subsequently,
a revision of this scenario was put forward \cite{RS2}, in which there was
only one 3-brane, of positive tension, essentially obtained from
the first scenario by sending the negative tension brane to the Cauchy
horizon of anti de Sitter space. The striking result found in this
second scenario is that, although the fifth dimension of space-time is now
infinite, the effective gravity theory on the single remaining 3-brane
nonetheless has $D=4$ and not $D=5$ leading behavior. The gravitational
potential for static sources starts out with a Newtonian $1/r$, corrected by
terms of order $\Lambda^{-1}/r^3$, where $\Lambda$ is the $D=5$ cosmological
constant. This ``binding of gravity'' to the 3-brane happens when a $D=5$
spacetime has a warped product structure, with the warp factor, \ie the factor
multiplying the $D=4$ submetric, decreasing as one recedes on either side from
the single Randall-Sundrum 3-brane. This corresponds in general terms to the
3-brane acting as a positive-tension source on the right-hand side of
the Einstein equations. It was not clear, however, whether this scenario
could arise from an explicit solution to a supergravity theory.

Links between the Randall-Sundrum model and supergravity were made in
Refs.\ \cite{Youm:2000bz,Youm:2000zp,Cvetic:1999fe,Cvetic:2000dz,
deAlwis:2000qc,deAlwis:2000pr}.
In \cite{Cvetic:1999fe,Cvetic:2000dz}, the $D=5$ 3-brane solutions to
the type II theory presented in \cite{Bremer} were used to make an
analogy to the Randall-Sundrum model. The explicit relation between
this construction and the specific Randall-Sundrum model was not
fully pinned down, however. This perspective was further elaborated in
\cite{deAlwis:2000qc,deAlwis:2000pr}.  Despite the existence of these
works, there still seems to be some confusion in the literature
as to whether the Randall-Sundrum model can in fact be obtained from
an explicit supergravity solution%
\footnote{The equivalence of the 
graviton propagator calculated from closed loops of the $N=4$ SCFT in 
the Maldacena picture and that calculated from tree graphs in the 
Randall-Sundrum picture was already strongly indicative of a supersymmetric 
Randall-Sundrum brane-world \cite{Duffliu}.}.
Moreover, there are powerful general arguments
\cite{Kallosh:2000tj,BeCvII,Ceresole:2000jd} as to why smooth supersymmetric
solutions obtained from $D=5$ gauged supergravity coupled to various
combinations of $D=5$ matter cannot reproduce a Randall-Sundrum scenario with
binding of gravity to the 3-brane. A key word here is `smooth.'  Although one
might well like to replace the Randall-Sundrum scenario, with its delta-function
source, by a smooth solution, experience with domain walls in supergravity
(\ie codimension-one brane solutions) shows them always to be based
upon a linear harmonic function in the $d=1$ codimension. In order for
such a solution to have a localized energy concentration, \ie a `brane,'
some kind of `kink' must be introduced into the linear harmonic function
so as to give a location to the domain wall. Thus, the search for a
smooth codimension one solution looks rather unlikely to be 
successful.%
\footnote{Some rigorous results along these lines have 
recently been spelled in Ref.\ \cite{maldacena}.  See also
Ref.\ \cite{Wijnholt:1999vk}.}
Moreover, the remainder of the argument of Refs.\ \cite{Kallosh:2000tj,BeCvII}
concerns the general behavior of renormalization group flows between
critical points of coupled supergravity-matter potentials. This gives the
impression that even if one were to relax the requirement of smoothness,
there would be no solution leading to the binding of gravity to the 3-brane.

In this paper, we shall first explicitly obtain the original (kinked)
Randall-Sundrum geometry from type IIB supergravity. This follows
from the work of Refs.\ \cite{Bremer,Cvetic:1999fe,deAlwis:2000pr}. This
construction makes essential use \cite{Cvetic:1999fe} of the `breathing mode'
of the $S^5$ dimensional reduction of type IIB supergravity of
Ref.\ \cite{Bremer}.
We shall show why this massive mode escapes the constraints on supersymmetric
flows by reason of its transforming in a representation
with AdS lowest energy $E_0=8>4$, thus falling outside the scope of the
analysis of Ref.\ \cite{Kallosh:2000tj,BeCvII}. 
The breathing-mode solutions, although Kaluza-Klein consistent in a
purely bosonic context containing just the breathing mode and gravity, do not
really correspond to a pure $D=5$ supergravity theory. The construction
retains an essential memory of its $D=10$ type IIB origin. This is particularly
so when one considers the superpartners of the breathing mode, which include
massive spin two modes that cannot be retained in a consistent truncation to a
finite number of $D=5$ fields.

Another memory of $D=10$ supergravity in the supersymmetric realization of the
Randall-Sundrum geometry resides in the $Z_2$ symmetry of this geometry. This
geometry is very similar to the $Z_2$ symmetric configuration of two M--theory
3-branes in $D=5$ that explicitly realizes the Ho\v{r}ava-Witten construction as
an M-theory brane solution \cite{losw1,losw2}. In the M--theory solution, the
$Z_2$ symmetry is central to the appearance of the orbifold, and it also plays a
critical r\^ole in the preservation of unbroken $D=4$ supersymmetry on the brane
world-volumes \cite{losw2}. The same is true in the double 3-brane type IIB
solution that we present as the supergravity realization of the Randall-Sundrum
geometry: continuity of the unbroken supersymmetry Killing spinor depends on the
way the $Z_2$ symmetry is implemented. In particular, in the M--theory case
\cite{losw2} as well as in the type IIB construction \cite{Kraus:1999it}, the
constant parameter determining the flux of the relevant underlying form field is
$Z_2$ odd, and so flips sign upon crossing either of the 3-branes; this flip is
crucial for the continuity of the unbroken supersymmetry parameter. Accordingly,
in the type IIB case as in the M-theory case, the $D=5$ theory is really
obtained from a dimensional reduction on a {\it pair} of Kaluza-Klein ans\"atze,
one on each side of the $Z_2$ symmetric spacetime. Although this construction
requires the presence of brane sources for the form-field flux, it is natural in
the context of the higher-dimensional M-- or type IIB theory. This split ansatz,
however, means that it is much less natural to view the geometry as 
arising in a single $D=5$ theory.

Having shown how to obtain the Randall-Sundrum model from type IIB supergravity,
we next set out to study the brane-wave oscillations of the solution.  This
analysis is quite natural in the type IIB analogue of the M--theory $Z_2$
symmetric double 3-brane construction \cite{losw1,losw2}. Although, as in
\cite{RS1}, this configuration involves both a positive and a negative tension
brane, thus leading to concerns about negative energies, we show that there
is a `mode-locking' phenomenon that reduces the zero-modes to just one
(positive energy) $D=4$, $N=4$ Maxwell multiplet in the case of one singly
charged brane. This happens because the $Z_2$-odd modes
turn out to be non-zero modes constrained to be related to Kaluza-Klein massive
modes by the Bianchi identities for the type IIB 5-form field strength
$H_{[5]}$ and for the gravitational curvature. Thus, one does not have
to make an explicit projection by hand into a $Z_2$-invariant subspace
of fields: this projection happens spontaneously, by normal Kaluza-Klein
dynamical mechanisms freezing out massive Kaluza-Klein modes. The type
IIB models considered here have the great advantage that one can carry
out more of the Kaluza-Klein analysis explicitly than in the analogous
discussion of M--theory reduced on Calabi-Yau spaces \cite{losw1,losw2}. But it
is to be expected that an analogous mode-locking mechanism will operate there as
well. And in that case, the mode-locking can be expected to lead to a
spontaneous appearance of $D=4$ chirality, thus generalizing the appearance of
chirality by explicit $Z_2$ projection.

\section{Supersymmetric domain walls and renormalization group flows}

While there are many ways of representing a metric on anti-de Sitter
space, perhaps the most natural form of the metric from a domain wall
point of view is given in terms of Poincar\'e coordinates
\begin{equation}
\label{eq:pads}
ds^2 = e^{-2gy} \eta_{\mu\nu}dx^\mu dx^\nu + dy^2.
\end{equation}
Written in this manner, the Minkowski signature boundary of AdS is
reached when $y\to-\infty$, while the point $y\to\infty$ is instead a
null surface, the AdS Killing horizon.  In the AdS/CFT correspondence,
this metric is viewed as the near-horizon geometry of $N$ coincident
D3-branes, which is described by ${\cal N}=4$ super Yang-Mills living on
the boundary.  Furthermore, the distance to the boundary is regarded as
an energy; from the bulk point of view $y\to-\infty$ is a flow to the UV,
while $y\to\infty$ is a flow to the IR.

The Randall-Sundrum brane-world is obtained by taking two Poincar\'e
patches of AdS, both given by (\ref{eq:pads}), and joining them at the
brane location $y=0$.  The resulting Randall-Sundrum metric has the form
\begin{equation}
\label{eq:rsads}
ds^2 = e^{-2g|y|} \eta_{\mu\nu}dx^\mu dx^\nu + dy^2,
\end{equation}
and its geometry gives rise to a localized graviton on the `Planck' brane.  
Presented as `an alternative to compactification', much has been
made of the fact that this scenario binds gravity even though the $y$
direction has an infinite extent.  Nevertheless, it is apparent from the
form of (\ref{eq:rsads}) that the Planck brane only lives in a tiny
portion of AdS, and that movement away from the brane flows towards the
Killing horizon and not towards the Minkowski boundary of AdS.  Had one
instead chosen to join together the $y<0$ regions of (\ref{eq:pads}),
the resulting geometry would preserve the vast majority of the original
space, including all of the portion of AdS near the boundary.  This then
would yield a divergent `localization' volume and give rise to a brane
of opposite character to the Randall-Sundrum brane, namely one that does
not bind gravity.

In fact, the above observation motivated the authors of
Ref.~\cite{Chan:2000ms} to view the Randall-Sundrum geometry as a warped
compactification of F-theory on a Calabi-Yau four-fold.  In this
picture, the warped geometry arises from the presence of D3-branes
situated on the elliptically fibered Calabi-Yau manifold.  The five
dimensional Randall-Sundrum universe then corresponds to the noncompact
four-dimensional spacetime with the addition of a single $y$ coordinate
which provides a preferred slicing of the internal space along flows
between separated stacks of D3-branes.  One thus sees that the
Randall-Sundrum brane itself is not identified with any one of the
D3-branes, but is instead viewed as an effective geometry that arises in
interpolating between the near-horizon locations of the D3-branes.  In
terms of the parametrization in (\ref{eq:rsads}), the D3-branes are located
at the horizons, $y=\pm\infty$, and the apparent infinite extent of the
$y$ coordinate is simply a result of the warping of the compact space by
the D3-branes themselves.  The localization of gravity is then explained
by the compactness of the underlying F-theory construction.  Heterotic
and M-theory realizations based on warped Calabi-Yau compactifications
have been examined in Ref.~\cite{Mayr:2000zd}.

Returning to a five-dimensional picture, there have been many
attempts to explain the Randall-Sundrum scenario from a supersymmetric
domain-wall point of view.  The advantage of this approach is that one
can generally ignore the added complications of the compactification of
the underlying IIB theory, and instead focus only on brane constructions
in the resulting $D=5$ gauged supergravity theory.  However, as we emphasize
below, it is important to realize that there is no reason (other than
simplicity) to expect that the relevant degrees of freedom lie only 
in the massless supergravity sector.  In fact, as emphasized in
\cite{Kallosh:2000tj,BeCvII}, massless gauged supergravity precludes the
localization of gravity on a brane. Thus massive fields are a
necessity.

For the Randall-Sundrum picture to be realistic, where the Planck brane
is a dynamical object, it would have to be supported by bulk scalar fields.
Thus, in the language of bulk renormalization group flow, we seek a
brane solution with stable flows to AdS critical points in the IR on both
sides of it.  This approach has been studied extensively in both the
AdS/CFT \cite{Skenderis:1999mm,Freedman:1999gp} and brane-world
\cite{Behrndt:2000kz,DeWolfe:1999cp,Kallosh:2000tj,BeCvII} pictures, with
considerable overlap.
Nevertheless, the distinction between flows of massless and massive
scalars has not always been made clear, so we wish to do so below.

Since we demand that the flow away from the brane is towards an AdS
background, the scalars must reach some fixed values corresponding to a
critical point in the potential.  Then, independent of any specific
model, at that point, we may expand the scalars about their fixed
values.  However before doing so, it is worth realizing that
representations in AdS differ from those in a flat background.

Recall that, for AdS$_5$, general representations of $SU(2,2)$ may be
labeled by $D(E_0,j_1,j_2)$ where $E_0$ is the lowest energy
(which may be given in terms of the natural mass scale of the AdS
background).  For scalars, $D(E_0,0,0)$, unitarity requires $E_0\ge1$
with $E_0=1$ corresponding to the singleton representation.  General
unitarity bounds for $SU(2,2)$ as well as for the $SU(2,2|N/2)$
superalgebras have been obtained in
\cite{Flato:1984te,Dobrev:1985qv,Bars:1983ep,Gunaydin:1988hb,Gunaydin:1999jc,%
Gunaydin:1998sw} (see also \cite{Ferrara}).  For a scalar
field in AdS$_5$, the mass is given in terms of $E_0$ by
$m^2=E_0(E_0-4)$, so that `massless' scalars in fact correspond to
$E_0=4$.  Of course, negative mass squared is not to be feared in an AdS
background, provided the Breitenlohner-Freedman bound
\cite{Breitenlohner:1982jf} is satisfied.  For this case it corresponds
to $m^2\ge -4$, which is saturated for $E_0=2$.

To be specific, we now consider the case of a brane supported by a single
scalar coupled to gravity, where the Lagrangian takes the form
\begin{equation}
\label{eq:lag}
e^{-1}{\cal L}= R-\half\partial\phi^2-V(\phi).
\end{equation}
While one may generalize by including more scalars, this single scalar
example is sufficient to bring out our conclusion.  The resulting equations
of motion have the form
\begin{eqnarray}
\label{eq:eom}
R_{MN}&=&\half\partial_M\phi\partial_N\phi+\ft13g_{MN}V(\phi),
\nonumber\\
\nabla^2\phi&=&\partial_\phi V(\phi).
\end{eqnarray}
Note that we do not insist that (\ref{eq:lag}) necessarily originates from a
supersymmetric theory. However in many cases we are of course interested in
supersymmetry.  This suggests the identification of a putative `superpotential'
$W(\phi)$ with 
\begin{equation}
\label{eq:vwrel}
V=(\partial_\phi W)^2-\fft23W^2,
\end{equation}
and the putative `transformations'
\begin{eqnarray}
\label{eq:susyx}
\delta\psi_\mu&=&[\nabla_\mu-\ft1{6\sqrt2}W\gamma_\mu]\epsilon,
\nonumber\\
\delta\lambda&=&\half[\gamma\cdot\partial\phi+\sqrt{2}\partial_\phi W]
\epsilon.
\end{eqnarray}
Identification of the above transformations with those of an actual
supergravity theory requires some care.%
\footnote{In this paper, we have not
performed a full Kaluza-Klein reduction of the type IIB supersymmetry
transformations, so all discussions of supersymmetry here are somewhat
schematic. They will be sufficient, however, to determine the necessary $Z_2$
behavior for having an unbroken $D=5$ supersymmetry.} In particular, from an
$N=2$ point of view,%
\footnote{We take $N=2$ to label minimal supersymmetry in $D=5$,
corresponding to $D=4$, $N=2$ supersymmetry.}
the field $\phi$ may reside in either a vector, tensor or hypermatter
multiplet, with possibly different forms of coupling to the fermions.
In all cases, the fields $(g_{\mu\nu},\psi_\mu)$ and $(\phi,\lambda)$ would
be part of a (not necessarily consistent) truncation of the actual
supergravity theory.  

As emphasized previously in discussions of holographic renormalization
group flows, the equations of motion following from a domain-wall ansatz
take on a simple form.  Starting with the metric
\begin{equation}
\label{eq:abans}
ds^2=e^{2A(y)}\eta_{\mu\nu}dx^\mu dx^\nu+e^{2B(y)}dy^2
\end{equation}
one obtains the following equations
\begin{eqnarray}
\label{eq:abpeom}
&&A'^2=\ft1{24}\phi'^2-\ft1{12}e^{2B}V,\nonumber\\
&&A''-A'B'=-\ft16\phi'^2,\nonumber\\
&&\phi''+(4A'-B')\phi'=e^{2B}\partial_\phi V,
\end{eqnarray}
where primes denote $y$ derivatives.  The first two equations were
obtained by combining components of the Einstein equation.  Note that
the three equations are not all independent, and we find it convenient
to focus only on the first two.

In codimension one, the second metric factor $e^{2B}$ is redundant, and
may be removed by defining a new coordinate $\tilde y=\int e^B dy$
(keeping in mind that explicit domain wall solutions often have a
simpler form when presented in terms of a metric with the $e^{2B}$
factor).  We proceed by setting $B=0$, so the equations resulting
from (\ref{eq:abpeom}) take the form
\begin{equation}
\label{eq:aflow}
A''=-\ft16\phi'^2,\qquad
A'^2=\ft1{24}\phi'^2-\ft1{12}V,\qquad
\phi''+4A'\phi'=\partial_\phi V.
\end{equation}
As emphasized in \cite{Freedman:1999gp,Skenderis:1999mm}, the first of
these equations indicates that $A''\le0$, with saturation of the
inequality corresponding to sitting in the pure AdS vacuum.  For the
present case, this has the consequence that the function $A(y)$ must be
concave-down, which is in fact exactly what is needed to support a
`kink-down' (\ie positive tension) Randall-Sundrum brane of the form 
(\ref{eq:rsads}) with a continuous metric function.

To study the behavior of the flow to the IR fixed point, we may expand
about the fixed value, $\phi_*$, of the scalar.  To quadratic order,
the potential then has the form
\begin{equation}
\label{eq:potex}
V=-12g^2+\half m^2(\phi-\phi_*)^2+\cdots,
\end{equation}
where the constant factor is chosen to give the conventional
normalization of the AdS curvature,
\begin{equation}
R_{MNPQ} = - g^2(g_{MP}g_{NQ}-g_{MQ}g_{NP}).
\end{equation}
While in some cases $g$ may coincide with the coupling constant of gauged
supergravity, we only take it to parameterize the AdS background at the
specific fixed point we are interested in.

We now insert (\ref{eq:potex})
into the second equation of (\ref{eq:aflow}) to find that $A(y)\approx\pm gy$,
at least up to linear order in $\phi$.  Thus we recover the expected linear
behavior giving rise to an AdS background.  Continuing with the $\phi$
equation of motion, and again working to linear order in $\phi$ (which
amounts to making the substitution $A'\approx\pm g$), we find
\begin{equation}
\label{eq:phieom}
\phi''\pm4g\phi'-m^2\phi\approx0,
\end{equation}
which has in general two solutions,
\begin{equation}
\label{eq:phisol}
\phi\approx\phi_*+c e^{-E_0A(y)},\quad\hbox{and}\quad
\phi\approx\phi_*+c e^{-(4-E_0)A(y)},
\end{equation}
where $E_0=2+\sqrt{(m/g)^2+4}\ge2$ is given exactly by the mass/$E_0$
relation for a scalar field in AdS space.  Additionally, for either flow,
the metric function behaves like
\begin{equation}
A\approx\pm gy-\ft1{24}(\phi-\phi_*)^2.
\end{equation}
Finally, this allows us to examine the IR flow, corresponding the to behavior
in the direction $A\to-\infty$.  We see that IR stability is ensured for
$E_0>4$ by taking the second solution of (\ref{eq:phisol}), while the flow
is always unstable for $2\le E_0<4$, and the massless case, $E_0=4$, is
marginal.

As a result, the above analysis indicates that $E_0>4$ is a necessary
condition for IR stability, and hence for the construction of a
Randall-Sundrum brane.  Note, furthermore, that this result was derived
without having to appeal to supersymmetry.  Thus it holds in general for
both BPS and non-BPS flows.  However, as we see below, BPS flows impose
a further condition on the relative signs of the terms in the superpotential.
This powerful and completely general result was in fact present, although
hidden in the discussion of Ref.~\cite{Kallosh:2000tj,BeCvII}.  However, in
\cite{Kallosh:2000tj,BeCvII}, only scalars residing in massless vector
multiplets
of $N=2$ gauged supergravity (\ie the ${\cal D}(2,0,0,0)$ representation,
where the last value denotes the $U(1)_r$ charge) were considered.  In
particular, the authors of \cite{Kallosh:2000tj} relied on the relation
$(\partial_i\partial_j W)_{cr} = \fft13 g_{ij}W_{cr}$ (in our normalization)
arising from very special geometry.  Such scalars always have $E_0=2$,
yielding the negative reported result.  Curiously, while it may not have
been appreciated that scalars in the decomposition of an $N=8$ gauged
supergravity multiplet reside in $N=2$ tensor and hypermatter multiplets
as well as vector multiplets, such $N=8$ scalars all have $E_0=2$, 3 or 4
so that they also do not lead to IR stable branes.

Turning now to the case of a supersymmetric flow, it is straightforward to
see from (\ref{eq:susyx}) that the Killing spinor conditions yield the first
order equations
\begin{eqnarray}
\label{eq:kse}
A'&=&\pm\fft1{3\sqrt{2}}e^BW\nonumber\\
\phi'&=&\mp\sqrt{2}e^B\partial_\phi W,
\end{eqnarray}
for a domain wall preserving exactly half of the supersymmetries.
This result may in fact also be derived without using the transformations
(\ref{eq:susyx}), but instead by a traditional BPS argument for finding
static minimum energy configurations \cite{DeWolfe:1999cp,Skenderis:1999mm}.
Combining both equations gives rise to a holographic renormalization
group flow
\begin{equation}
\label{eq:rgfl}
{d\phi\over dA} = -6{\partial_\phi W\over W},
\end{equation}
consistent with the second order equations (\ref{eq:abpeom}).  In this case,
we expand the superpotential as
\begin{equation}
\label{eq:suppot}
W=\pm3\sqrt{2}g(1+\ft1{12}\lambda(\phi-\phi_*)^2+\cdots)
\end{equation}
corresponding to the potential (\ref{eq:potex}), provided $\lambda$ is
identified with either $E_0$ or $4-E_0$.  Note that this introduces a
two-fold ambiguity.  However this is in fact somewhat artificial, since
knowledge of the actual supersymmetric theory would fully determine the
superpotential (but see {\it e.g.}\ Ref.~\cite{deBoer:1999xf} for a
discussion on the relation between $V$ and $W$ without supersymmetry).
In contrast to (\ref{eq:phisol}), the supersymmetric flow condition,
(\ref{eq:rgfl}), gives only a single approach to the fixed point
\begin{equation}
\phi\approx\phi_*+ce^{-\lambda A(y)}.
\end{equation}
As a result, for a BPS flow, not only do we require $E_0>4$, but also we
learn from the above analysis that $\lambda=4-E_0$ must be negative
in the superpotential, (\ref{eq:suppot}).  The requirement of $\lambda<0$
was previously noted in \cite{Kallosh:2000tj}.

This connection between $E_0$ and the behavior of a scalar field in AdS
was initially made in investigations of the Maldacena conjecture
\cite{Gubser:1998bc,Witten:1998qj}, where $E_0$ was related to the
conformal dimension of appropriate operators on the CFT side of the AdS/CFT
conjecture.  In this case, (\ref{eq:phieom}) taken with exact equality is
simply the massive scalar equation in the reference AdS background
(\ref{eq:pads}).  This in itself highlights the similarity between the
brane-world scenario and the AdS/CFT conjecture.  In some sense, the
Randall-Sundrum brane, being inserted at some fixed location in AdS, cuts
off the flow to the UV and hence may be described by a Maldacena CFT cut off
at some energy scale related to the location of the brane.

\section{Breathing mode domain walls and the
brane-world\label{sec:breathingwalls}}

Based on the preceding analysis, it is clear that consideration of the
massless sector of ($N=2$, 4 or 8) gauged supergravities alone does not lead
to realistic brane-world configurations.  However, for a five dimensional
model originating from IIB theory, many other degrees of freedom may come
into play.  While round $S^5$ compactifications of IIB supergravity yield
$N=8$ gauged supergravity at the massless level, this is also accompanied
by a Kaluza-Klein tower of massive states.  In general, consistent
truncations of sphere reductions are a delicate matter
\cite{kksugra,Cvetic:2000dm}.
However it is consistent to include the breathing mode $\varphi$ in the
truncation: although it lives in a massive supermultiplet, it is
nevertheless a gauge singlet.%
\footnote{Note that we use $\varphi$ to denote the breathing mode rather than
$\phi$, in order to emphasize that it is distinct from the $D=10$ dilaton of the
type IIB theory.}
Domain walls supported by the breathing mode have been investigated in
Refs.\ \cite{Bremer,Cvetic:1999fe,deAlwis:2000pr}, and have recently been
suggested as possible realizations of the brane-world scenario.

To make connection with the Randall-Sundrum model, we examine type IIB
string theory compactified on $S^5$.  This sphere reduction, with the
inclusion of a single squashing mode along with the breathing mode,
was investigated in \cite{Bremer}.  Focusing only on the scalar modes,
the resulting five dimensional Lagrangian is
\begin{equation}
e^{-1}{\cal L}_5= R-\half\partial\tilde\varphi^2-\half\partial\tilde f^2
-V(\tilde\varphi,\tilde f).
\end{equation}
The scalar potential has the form
\begin{equation}
V(\tilde\varphi,\tilde f)=8m^2e^{{10\over\sqrt{15}}\tilde\varphi}
+e^{{4\over\sqrt{15}}\tilde\varphi}(\mu^2e^{{6\over\sqrt{10}}\tilde f}
-R_4 e^{{1\over\sqrt{10}}\tilde f}),
\end{equation}
where the constants $(m,\mu,R_4)$ are parameters of the compactification
\cite{Bremer}.

While this potential may now be expanded in the form of
Eqn.~(\ref{eq:potex}), it is perhaps more enlightening to first express
it in the form of a `superpotential' according to (\ref{eq:vwrel}).  We
find
\begin{equation}
W=2\sqrt{2}m
e^{{5\over\sqrt{15}}\tilde\varphi}
-e^{{2\over\sqrt{15}}\tilde\varphi}(\sqrt{2}\mu e^{{3\over\sqrt{10}}\tilde f}
+{R_4\over2\sqrt{2}\mu} e^{-{2\over\sqrt{10}}\tilde f}).
\label{eq:sppot}
\end{equation}
Note that there is a slight sign ambiguity in inverting (\ref{eq:vwrel});
here we have chosen the signs so that $W$ has a critical point at
\begin{equation}
e^{{3\over\sqrt{15}}\tilde\varphi_*}=
{\mu\over2m}\left(R_4\over6\mu^2\right)^{3\over5},\qquad
e^{{5\over\sqrt{10}}\tilde f_*}={R_4\over 6\mu^2},
\end{equation}
corresponding to that of $V$ as well.  Expansion of $W$ then gives
\begin{equation}
W=-3\sqrt{2}m\left(\mu\over2m\right)^{5\over3}
\left(R_4\over6\mu^2\right)[1-\ft13(\tilde\varphi-\tilde\varphi_*)^2
+\half(\tilde f-\tilde f_*)^2+\cdots].
\end{equation}
Comparison with (\ref{eq:suppot}) then demonstrates explicitly that
the breathing mode $\tilde\varphi$ has $E_0=8$ while the squashing mode
$\tilde f$ has $E_0=6$.  Curiously, the two modes enter with opposite
signs in $W$.  While this $N=8$ symmetric critical point is indeed a
minimum of the potential, it is only a saddle point of $W$.

The consequences for the resulting supersymmetric flow were
investigated in the previous section.  For supersymmetric flows, this
critical point is IR stable for the breathing mode, while it is unstable
for the squashing mode.  This indicates explicitly that simply having a
domain wall supported by a scalar with $E_0>4$ may be insufficient to
ensure the stability of a supersymmetric Randall-Sundrum configuration.
Nevertheless, we have now seen why use of the massive breathing mode of
sphere reductions has been successful in constructing brane-world
domain walls \cite{Cvetic:1999fe,deAlwis:2000pr}, avoiding the
limitations on supersymmetric flows presented in
Ref.\ \cite{Kallosh:2000tj,BeCvII}.

To proceed, we now truncate out the squashing mode by setting $\tilde f=0$
and $R_4=6\mu^2=\fft65R_5$.  After dropping tildes, the resulting potential
for the breathing mode is simply
\begin{equation}
\label{eq:5pot}
V(\varphi)=8m^2e^{{10\over\sqrt{15}}\varphi} -R_5e^{{4\over\sqrt{15}}\varphi},
\end{equation}
and has an AdS minimum at
\begin{equation}
e^{{6\over\sqrt{15}}\varphi_*}={R_5\over20m^2}.
\end{equation}
Here $R_5$ is the curvature scalar of the round $S^5$, arising from the
type IIB Kaluza-Klein ansatz \cite{Bremer}
\begin{eqnarray}
\label{eq:kka}
ds_{10}^2&=&e^{2\alpha\varphi}ds_5^2+e^{2\beta\varphi}ds^2(S^5)\nonumber\\
H_{[5]}&=&4me^{8\alpha\varphi}\epsilon_{[5]}+4m\epsilon_{[5]}(S^5),
\end{eqnarray}
where
\begin{equation}
\alpha=\fft14\sqrt{\fft53},\qquad\beta=-\fft35\alpha.
\end{equation}
This also indicates that $m$ is essentially the 5-form flux of the
Freund-Rubin compactification.  Thus the two parameters $m$ and $R_5$ of
the five dimensional potential, (\ref{eq:5pot}), have their origin in the
Kaluza-Klein compactification from ten dimensions. Note that the Kaluza-Klein
ansatz (\ref{eq:kka}) for the $H_{[5]}$ field strength implies that the
Freund-Rubin parameter $m$ must be odd under transformations $y\rightarrow -y$.
In order for this to be realized as a symmetry of the type IIB theory, this
`lower' $D=5$ transformation must be accompanied by an orientation-reversing
transformation of $S^5$, so that the self-dual structure of $H_{[5]}$ is
preserved, but with $m\rightarrow -m$.  By the `skew-whiffing theorem'
\cite{kksugra}, both orientations have the same (maximal) supersymmetry in the
case of $S^5$.  For any other compactifying 5-manifold the
supersymmetries would not match.

For a complete truncation of the sphere compactification down to $D=5$, in which
all Kaluza-Klein modes except for the breathing mode are discarded, the two
parameters $m$ and $R_5$ satisfy trivial Bianchi identities, and hence must
be constant.  In this case only a single combination of the two is actually
physical.  The constant parameter $R_5$ may then be viewed as a necessary
dimensionful parameter for measuring coordinate distances on the five sphere
(much as one would have to introduce a length scale $L$ for toroidal
compactification, where periodic coordinates are identified as $y=y+2\pi L$).
The actual invariant (physical) size of the five sphere is then set by the
expectation of the breathing mode $\varphi$.
To see formally how $R_5$ may be scaled away, consider a shift of $\varphi$
along with a scaling of $m$
\begin{equation}
\label{eq:lscal}
\varphi\to\varphi+{\sqrt{15}\over4}\log\lambda,\qquad
m\to m\lambda^{-{5\over4}}.
\end{equation}
This transformation has the effect of multiplying $R_5$ by $\lambda$ in the
potential (\ref{eq:5pot}), so that an appropriate choice of $\lambda$ may be
used to scale $R_5$ to any desired value.  A particularly natural choice
would be to set $R_5=20m^2$, so that the AdS critical point is reached at
$\varphi_*=0$.  From a ten dimensional point of view, the transformation
(\ref{eq:lscal}) results in
\begin{eqnarray}
ds_{10}^2&=&\lambda^{5\over8}[e^{2\alpha\varphi}ds_5^2+e^{2\beta\varphi}
\lambda^{-1}ds^2(S^5)]\nonumber\\
H_{[5]}&=&\lambda^{5\over2}[4me^{8\alpha\varphi}\epsilon_{[5]}
+4m\lambda^{-{5\over2}}\epsilon_{[5]}(S^5)],
\end{eqnarray}
which is thus a rescaling of $S^5$ combined with a $D=5$ `trombone'
symmetry.

However, as we will discuss in the following Section, if one no longer
truncates out the additional Kaluza-Klein modes, then both $m$ and
$R_5$ no longer need to be taken as constant.  In this case, attempts
to scale away $R_5(x)$ would result in a dynamical scaling by $\lambda(x)$.
In this sense one simply trades one parameter for another, and cannot fully
eliminate $R_5$.  With this in mind, we maintain both parameters $m$ and $R_5$
in the solution below.

Breathing mode domain wall solutions follow by making the standard ansatz
(\ref{eq:abans}) and by solving the resulting equations (\ref{eq:abpeom}).
As mentioned above, keeping two independent factors in the ansatz, $A(y)$
and $B(y)$, is redundant.  For $B=0$, the solution was presented in
\cite{Cvetic:1999fe}, while it was originally presented in \cite{Bremer}
with a different choice of coordinates.  The advantage of the original
choice is its highlighting of a linear harmonic function as a natural
feature of codimension one $p$-brane solutions.  This solution has the
basic form \cite{Bremer}
\begin{eqnarray}
\label{eq:brsol}
&&e^{-{7\over\sqrt{15}}\varphi}=H,\qquad
e^{4A}=e^{-B}=\tilde b_1H^{2\over7}+\tilde b_2 H^{5\over7},\nonumber\\
&&\qquad\qquad H=e^{-{7\over\sqrt{15}}\varphi_0}+ky,
\end{eqnarray}
where
\begin{equation}
\label{eq:bvals}
\tilde b_1=\eta_1{28m\over3|k|},\qquad
\tilde b_2=\eta_2{14\sqrt{5R_5}\over15|k|}.
\end{equation}
Here $\eta_{1,2}=\pm1$ are in general independent choices of signs for the
solution.  For our purposes they are fixed by requiring an appropriate AdS
limit for $\varphi\to\varphi_*$.  This gives $\eta_2=-\eta_1$ and $\eta_1$
chosen so that $e^{4A}>0$ in order for the metric to be real at a given initial value
of $y$.

The linear harmonic function $H$ is restricted to be nonnegative.
Examination of the solution indicates that the AdS horizon is located
at $H=H_*\equiv e^{-7\varphi_*/\sqrt{15}}$, where $e^{4A}$ vanishes.
For initial $H>H_*$ the five dimensional space asymptotically flattens out
as $H\to\infty$, with a corresponding limit for the scalar
field $\varphi\to-\infty$, yielding an asymptotically vanishing scalar
potential.  This case is the second branch of Ref.~\cite{Cvetic:1999fe}, where
it was referred to as a hybrid Type II and dilatonic domain wall.  On the other
hand, for initial $H<H_*$, the solution soon runs into a singularity at $H=0$.
Note, however, that if one starts with a solution with $H>H_*$ initially and
signs $\eta_{1,2}$ chosen so as to make  $e^{4A}>0$ initially, but then
follows the evolution of $H$ within the spacetime through the $H=H_*$
horizon, the metric in the region with $H<H_*$ becomes complex, so
one should really treat the region below the horizon using different,
appropriately chosen coordinates. Both the $H>H_*$ and $H<H_*$ cases
have a natural interpretation in the lifting of (\ref{eq:brsol}) to
ten dimensions.  In the IIB theory, (\ref{eq:brsol}) lifts directly
to the geometry of $N$ coincident D3-branes with total charge $\tilde
k=m(20/R_5)^{5/2}$ \cite{Bremer}.  The two regions $H\gtlt H_*$ then
correspond to the regions either `outside' or `inside' the D3-brane
horizon.  This furthermore demonstrates that the first, $H>H_*$, case is
nothing but the conventional near horizon limit occurring prominently
in the Maldacena conjecture.  The second, $H<H_*$, case is unphysical
as it stands, however, as it sees a different region of the D3-brane
geometry containing a singularity.

Neither case by itself provides a suitable framework for a Randall-Sundrum
configuration.  While in one direction one may reach an AdS horizon,
in the other direction one will either run into a singularity or on out
into unbounded flat space.  One obvious possibility for obtaining an
asymptotically AdS space on both sides of a domain wall is to reflect the
solution at $y=0$, imposing thus a $y\to-y$ $Z_2$ symmetry.  The resulting
two-sided domain wall, supported by an absolute value kink in the linear
harmonic function
\begin{equation}
\label{eq:ksol}
H=e^{-{7\over\sqrt{15}}\varphi_0}+k|y|,
\end{equation}
was in fact how the solution was originally presented in \cite{Bremer}.
The presence of such a kink is rather natural for a codimension one
object.  Supergravity $p$-brane solutions are generally supported by
$\delta$-function sources at the locations of the branes themselves,
and this remains true for domain walls.  Passing through a domain wall,
one jumps through a sheet of charge, and this jump in charge manifests
itself in a change in the slope of the linear harmonic function. A priori,
the slope could take any values on the two sides of the domain wall,
but clearly the $Z_2$ symmetric jump from $k$ to $-k$ is a natural
configuration. We shall see that this configuration is distinguished
also by preserving unbroken supersymmetry.

For either the plain unkinked (\ref{eq:brsol}) or the kinked
(\ref{eq:ksol}) solution, the slope $k$ may be scaled away by taking $y\to
y/|k|$ and $x^\mu \to x^\mu |k|^{1/4}$.  This explains why the apparent
domain wall charge $k$ is not directly related to lifted quantities
such as the D3-brane charge $\tilde k$.  However, note that this scaling
does not eliminate the sign of $k$, leaving thus a distinction between
the slope up and slope down possibilities.  For discussions of multiple
domain wall configurations or brane fluctuations, it is more convenient
to retain $k$.

If one chooses to restrict the coordinate $y$ in (\ref{eq:ksol}) to range
only over the interval $-y_0\le y\le y_0$, identifying the points $y_0$
and $-y_0$, then one obtains a $Z_2$ symmetric solution that can serve as the
background for a $Z_2$ orbifold construction. This orbifold construction is
analogous to the treatment of M--theory 3-branes given in \cite{losw1,losw2} as
a brane realization of the Ho\v{r}ava-Witten
$S^1/Z_2$ orbifold, and has also been proposed in the Randall-Sundrum context in
\cite{deAlwis:2000pr}.  The identification of $y_0$ and $-y_0$
essentially reproduces the original Randall-Sundrum model \cite{RS1}
with both an attractive and a repulsive brane (if one chooses $k<0$,
then the attractive brane is the one located at $y=0$).  From the
five-dimensional point of view, the $y\to-y$ $Z_2$ map is a parity flip. 
As we have mentioned above, however, this alone is not a good symmetry of the
underlying type IIB theory.  In order for this transformation to be
compatible with the round-sphere compactification of the IIB theory, this $Z_2$
transformation must combine the flip in $y$ with an orientation-reversing
transformation \cite{Kraus:1999it} of the $S^5$. For example, an allowable
transformation flips all six of the coordinates transverse to the underlying
$D=10$ D3-brane. The net effect is to send $m\to-m$ as well as $y\to-y$.

This orientation reversal has important consequences for the supersymmetry
transformations (\ref{eq:susyx}), since the superpotential $W$ also flips,
$W\to-W$, under these transformations. Actually, this is what one wants,
because if the superpotential were to not to flip in this way, then all
supersymmetries would be broken by the domain wall, and it would then no
longer be BPS.  To see this, consider for example the $\delta\lambda$
transformation for the solution (\ref{eq:brsol}) with the linear harmonic
function (\ref{eq:ksol}).  By truncating out the squashing mode from
(\ref{eq:sppot}), one arrives at the breathing mode superpotential:
\begin{equation}
W=\sqrt{2}m\Bigl[2e^{{5\over\sqrt{15}}\varphi}-5\sqrt{R_5\over20m^2}
e^{{2\over\sqrt{15}}\varphi}\Bigr].
\end{equation}
Written as above, this clearly changes sign as $m\to-m$.  On the other hand,
If one were to assume instead that $W$ remains invariant, one would find
\begin{eqnarray}
\partial_\varphi W&=&\ft23\sqrt{30}m\Bigl[
e^{{5\over\sqrt{15}}\varphi}-\sqrt{R_5\over20m^2}
e^{{2\over\sqrt{15}}\varphi}\Bigr]\nonumber\\
&=&{\sqrt{30}\over14}|k|H^{-1}
(|\tilde b_1| H^{2\over7}-|\tilde b_2| H^{5\over7})\nonumber\\
&=&-{1\over\sqrt{2}}|\varphi'|e^{-B},
\end{eqnarray}
where the signs $\eta_{1,2}$ have been chosen to obtain the outside
(\ie $H>H_*$) AdS solution.  Inserting this into (\ref{eq:susyx}) we
would find
\begin{equation}
\label{eq:badsusy}
\delta\lambda=\half e^{-B}(\gamma^{\overline{y}}\varphi'-|\varphi'|)\epsilon\ ,
\end{equation}
where $\bar y$ denotes a local Lorentz index.
Because of the absolute value in the linear harmonic function
(\ref{eq:ksol}), $\varphi'$ changes sign on opposite sides of $y=0$.
Therefore the assumption of an invariant $W$ would leave no possibility of
obtaining a Killing spinor that is consistently defined on both sides of
$y=0$.  If one were to attempt to patch together separate Killing spinors
on both sides of $y=0$, in the case of an invariant $W$, the $y\gtlt0$
projections on the supersymmetry parameter would be into mutually
orthogonal components,
$(1+\gamma^{\overline{y}})\epsilon_+=0$ versus
$(1-\gamma^{\overline{y}})\epsilon_-=0$.  However, since the superpotential
{\it does} change sign under the $Z_2$, the absolute value in
(\ref{eq:badsusy}) is in fact not present, and we accordingly find global
Killing spinors of the form
$\epsilon = e^{A/2}(1+\gamma^{\overline{y}})\epsilon_0$. Similar
considerations apply at the location of the second kink in the $Z_2$
invariant background. If one expands the theory in modes about this
$Z_2$ invariant background, keeping only the $Z_2$ invariant modes, the
resulting theory is equivalent to one defined on an $S^1/Z_2$ orbifold.

As we have just demonstrated, the domain wall solution is always one half
supersymmetric, with or without the absolute value kink.  In particular,
the $Z_2$ orbifolding has not destroyed any further supersymmetry
beyond the original half-BPS solution.  On the other hand, there is
no restoration of supersymmetry either in the presence of of a kink.
Consider taking a simultaneous limit $k\to0$ and $\varphi_0\to\varphi_*$.
Without the kink, this limit would yield pure AdS, \ie the D3-brane near
horizon limit in which full supersymmetry is restored.  But with the kink,
one obtains instead a $Z_2$ symmetric patching of AdS, with a Randall-Sundrum
brane located, say, at $y=0$.  The presence of the orbifold fixed
point prevents the full supersymmetry from being restored. However,
this is fully expected when a domain wall is present.  Although the
$Z_2$ symmetrization introduces an absolute value into functions, the
Killing spinor equations are of first order, and so do not see any
$\delta$-function singularities.  As long as the conditions (\ref{eq:kse})
are satisfied, the solution remains supersymmetric.

Of course the second order equations of motion will see the $\delta$-function
brane source.  For the solution (\ref{eq:ksol}), we find that the extra
source terms at $y=0$ are
\begin{eqnarray}
\label{eq:bsource}
T_{MN}^{\rm brane}&=&
\hphantom{-4\sqrt{5\over3}}
{3k\over14}(2\tilde b_1^2e^{{3\over\sqrt{15}}\varphi_0}
+5\tilde b_2^2e^{-{3\over\sqrt{15}}\varphi_0} -7|\tilde b_1\tilde b_2|)
\delta(y)\delta_M^\mu\delta_N^{\nu\vphantom{\mu}}
g_{\mu\nu},\nonumber\\
Q^{\rm brane}&=&-4\sqrt{5\over3}{3k\over14}
(\tilde b_1^2e^{{3\over\sqrt{15}}\varphi_0}
+\tilde b_2^2e^{-{3\over\sqrt{15}}\varphi_0}-2|\tilde b_1\tilde b_2|)
\delta(y),
\end{eqnarray}
assuming $\tilde b_1 \tilde b_2 < 0$ as indicated above.
These `brane sources' enter in the equations of motion as
\begin{eqnarray}
R_{MN}-\half g_{MN}R&=&T^\varphi_{MN}+T_{MN}^{\rm brane},\nonumber\\
\nabla^2\varphi&=&\partial_\varphi V(\varphi)+Q^{\rm brane},
\end{eqnarray}
where
\begin{equation}
T_{MN} = \half(\partial_M\varphi\partial_N\varphi
-\half g_{MN}\partial\varphi^2)-\half g_{MN}V(\varphi).
\end{equation}
Depending on the sign of $k$, the branes have either positive or negative
energy density.  However, in both cases the relation between charge and
tension is the same, so the branes may be stacked up in BPS configurations.

We have thus seen that the kinks at the brane locations have different
consequences for supersymmetry and for the equations of motion.  Since the
supersymmetry variations and Killing spinor conditions are of first order,
the kinks give rise to possibly discontinuous quantities, but no
$\delta$-function singularities.  On the other hand, the equations of motion 
will be sensitive to the additional $\delta$-function sources.  Although one
may view the equations of motion as a composition of two supersymmetries,
there is no contradiction in the presence and absence of the
$\delta$-function terms since the Killing spinor equations only give rise to
a subset of the full equations of motion.  To see this consider again for
simplicity the $\delta\lambda$ transformation, (\ref{eq:susyx})
\begin{eqnarray}
\delta\lambda&=&\half[\gamma\cdot\partial_\phi+\sqrt{2}\partial_\varphi
W]\epsilon\nonumber\\
&=&\half e^{-B}[\varphi' \gamma^{\overline{y}}
+\sqrt{2}e^B\partial_\varphi W ]\epsilon.
\end{eqnarray}
Partial breaking of supersymmetry then demands the BPS condition%
\footnote{One may of course choose the other sign if so desired.  However
this is a global choice, and must be consistent in all patches of space.}
$\varphi'=-\sqrt{2}e^B\partial_\varphi W$, relating the scalar to its
potential.  Similarly, vanishing of the gravitino relates the metric to the
scalar potential, $A'=e^BW/3\sqrt{2}$, as given in (\ref{eq:kse}).
Now consider deriving the second order $A''$ equation of motion by taking a
derivative of $e^{-B}A'$
\begin{equation}
e^{-B}(A''-A'B') = {1\over3\sqrt{2}}W'
\end{equation}
For a continuous $W$, one simply uses the chain rule,
$W'=\partial_\varphi W\phi'$, and substitutes in the $\varphi'$ equation
to arrive at the $A''$ equation of motion given in (\ref{eq:abpeom}).
However, the assumption of a continuous $W$ is actually too strong.  For the
$Z_2$ invariant case, where $W$ changes sign at the brane (say at $y=0$),
one would also pick up a source term upon differentiating, resulting in
\begin{equation}
A''-A'B'=-\ft16\varphi'^2+\ft{\sqrt{2}}3e^BW\delta(y).
\end{equation}
Thus, while supersymmetry implies most of the equations of motion, it does
not in fact determine all of them.  In fact, for higher codimension
branes, there is even more slack between the BPS conditions and the
equations of motion.  The harmonic function condition, of primary importance
in brane constructions, is generally a consequence of the equations of
motion, and {\it not} supersymmetry \cite{Duff:1995an,Duff:1998me}.

\section{D3-branes and the world in ten dimensions}

Until now we have focused almost exclusively on the five-dimensional
viewpoint of the Randall-Sundrum scenario.  Since the breathing mode domain
wall has its origins in the $S^5$ compactification of IIB theory, it has
a natural interpretation in terms of IIB D3-branes \cite{Bremer}.  Following
this connection from the brane-world geometry to breathing mode branes
and then to D3-branes, one is led to a realization of the Randall-Sundrum
scenario in terms of IIB theory in an appropriate D3-brane background.

While the lifting of the breathing mode brane to patches of the D3 geometry
is straightforward, the resulting configuration has unusual features.
Following \cite{Bremer}, lifting of the solution given in (\ref{eq:brsol})
proceeds by identifying a ten-dimensional Schwarzschild coordinate
\begin{equation}
\label{eq:rhoy}
\rho=\sqrt{20\over R_5}H^{\fft3{28}}
\end{equation}
Using the charge relation $\tilde k=m(20/R_5)^{5/2}$ \cite{Bremer} and the
Kaluza-Klein ansatz (\ref{eq:kka}), one finds the resulting ten-dimensional
metric
\begin{equation}
ds_{10}^2=\tilde b_2^{\fft12}{\textstyle
\left(1-{\tilde k\over\rho^4}\right)^{\fft12}} dx_\mu^2 +
{\textstyle\left(1-{\tilde k\over\rho^4}\right)^{-2}} d\rho^2
+\rho^2d\Omega_5^2,
\end{equation}
which is that of $N$ D3-branes of total charge $\tilde k$
\cite{Horowitz:1991cd,Duff:1991pe}.  A further change
of coordinates, $r^4=\rho^4-\tilde k$, may be performed to transform this
into standard isotropic form
\begin{equation}
\label{eq:isomet}
ds_{10}^2=\sqrt{\tilde b_2}H_{D3}^{-1/2} dx_\mu^2 + H_{D3}^{1/2}(
dr^2 + r^2d\Omega_5^2),
\end{equation}
with a harmonic function $H_{D3}=1+\tilde k/r^4$.  Note that the constant
$\tilde b_2$ may easily be scaled out of the longitudinal coordinates.

For the $Z_2$ symmetric configuration, obtained by kinking the linear
harmonic function, (\ref{eq:ksol}), we see that $H$ is a double valued
function of $y$.  This has the consequence that the lifting relation
(\ref{eq:rhoy}) is similarly double valued; opposite sides of the
breathing-mode brane lift to identical $\rho$ values.  While the
orbifold picture corresponds to a single slice of the D3-brane geometry,
$\rho\in [\rho_-,\rho_+]$, the full circle compactification instead corresponds
to two copies of the D3-brane geometry patched together at $\rho_-$ and
$\rho_+$.  Note that the AdS horizon, located at $H_*$, lifts to the
D3-brane horizon, located at $\rho_*=\tilde k^{1/4}$.  Thus taking the
Randall-Sundrum configuration (kink down with $H>H_*$) and pushing the
second brane off to the Cauchy horizon corresponds in ten dimensions to
taking two copies of the near-horizon geometry of $N$ D3-branes, and gluing
them together at a value $\rho_0$ of the Schwarzschild coordinate
corresponding to the initial value $H_0$ of the linear harmonic function.

For this Randall-Sundrum configuration, it is instructive to `unfold' the 
doubled metric, (\ref{eq:isomet}), by defining a new radial coordinate
$\xi \in [-r_0,r_0]$ such that $r=r_0-|\xi|$.  After scaling out $\tilde
b_2$ from (\ref{eq:isomet}), the lifted Randall-Sundrum metric has the
form
\begin{equation}
ds_{10}^2={\textstyle\left(1+{\tilde k\over (r_0-|\xi|)^4}\right)^{-\fft12}}
dx_\mu^2 +{\textstyle\left(1+{\tilde k\over (r_0-|\xi|)^4}\right)^{\fft12}}
(d\xi^2+(r_0-|\xi|)^2d\Omega_5^2).
\end{equation}
The positive tension brane is located at $\xi=0$, while the negative
tension brane is pushed off to the AdS horizon at $\xi=\pm r_0$ (the two
values are identified under the $Z_2$ orbifolding).  As seen explicitly
here, this act of patching together two stacks of D3-branes essentially
compactifies the six-dimensional space transverse to the branes, and
also introduces a curvature discontinuity at $\xi=0$, the location of the
patching.  Furthermore, this compactification introduces a charge
conservation condition, implying that the net D3 charge must vanish.
Thus the resulting kink at $\xi=0$ must include a stack of $2N$
negative tension D3-branes, with $-2N$ units of charge soaking up
the $N+N$ units of charge from the two stacks of positive tension D3-branes.

The question arises, however, whether placing this stack of $2N$
negative tension D3-branes at $\xi=0$ is sufficient for generating the
kinked Randall-Sundrum geometry.  Furthermore, the reduction of D3-brane
tension from $D=10$ to $D=5$ yields the simple result
$T_{D=5}=T_{D=10}$.  In addition to giving rise to the tension discrepancy
pointed out in \cite{Kraus:1999it}, it also leaves unexplained how
positive $D=5$ tension arises from negative $D=10$ tension.  As it turns
out, the resolution to both issues is the realization that the $Z_2$
orbifolding, or the doubling of spacetime, itself gives rise to a
positive tension contribution at $\xi=0$, the location of the kink.  Of
course, it is easy to see that the net tension has to be positive, as that
is what is required to `fold up' or compactify the space transverse to the
branes.  The resulting picture is one of negative tension D3-branes
trapped on a positive tension $Z_2$ orbifold plane giving rise to a
composite description of the Randall-Sundrum configuration \cite{wip}.

By starting with a brane-world scenario on a circle, one obviously obtains a
compact Kaluza-Klein geometry, corresponding to expanding IIB theory about
a ${\cal M}^{1,3}\times S_1\times S^5$.  The $S_1$ coordinate $y$ lifts to
the radial coordinate $\rho$, living in a restricted annular range between the
two D3 source shells in a double D3-brane background.  Of course there is no
surprise in starting with a compact geometry and lifting it to another compact
scenario.  However, by taking the limit of placing the second brane at the
Cauchy horizon of AdS, one effectively decompactifies the original
Randall-Sundrum geometry of \cite{RS1} into the picture of \cite{RS2}. 
Nevertheless, from a ten-dimensional point of view, this corresponds to simply
extending the range of $\rho$ a finite distance so as to reach the doubled
D3-brane horizon: the internal space remains compact (at least if the
inside-horizon brane cores are disregarded).  By smoothing out the
patching of the double D3-brane configuration, one presumably obtains a warped
compactification with an internal six-manifold in the spirit of
\cite{Chan:2000ms}.

To complete this D3-brane picture of the brane-world,
we present the limit in which the $Z_2$
symmetric supergravity solution literally reproduces the Randall-Sundrum
configuration of a single positive-tension `kink-down' brane between
two patches of anti de Sitter space \cite{RS2}. Starting from the $D=5$
3-brane metric (\ref{eq:brsol}) with $\tilde b_2>0$, $\tilde b_1<0$,
$k<0$, we want to take a limit as $k\to 0_-$. However, the inverse power
of $k$ in $\tilde b_1$ and $\tilde b_2$ (\ref{eq:bvals}) makes this appear
singular. The cure for this is to take a coordinated limit as $k\to 0_-$
and $\varphi\to\varphi_*$. We implement this explicitly by taking
\begin{equation}
\label{eq:kzerolim}
e^{-{7\over\sqrt{15}}\varphi_0} = \left(20m^2\over R_5\right)^{7\over6}
+ \beta |k|\ ,\qquad \beta>0\ .
\end{equation}
Note that for $\beta>0$, one has
$e^{-{7\over\sqrt{15}}\varphi_0}>e^{-{7\over\sqrt{15}}\varphi_*}$,
\ie $H_0>H_*$.  Accordingly, for finite $k<0$, the harmonic function $H$
decreases from its value $H_0$, reaching the Cauchy horizon value $H_*$
at $y=y_h$. This is the natural point at which to make an identification
$y_h\leftrightarrow-y_h$, putting the second (negative tension)
3-brane at the
horizon. For finite $k$, one thus has a ``semi-interpolating soliton''
in the sense that one of the asymptotic limits of the solution, but not
both, corresponds to a vacuum solution of the theory, in this case the AdS
space with asymptotic scalar $\varphi_*$. At the Randall-Sundrum brane,
however, there is no horizon.

Taking the joint limit defined by (\ref{eq:kzerolim}) as $k\to 0_-$,
the difference between the two harmonic functions in $e^{2A}$ partially
cancels, giving an expression proportional to $k$, which cancels the
$k$ in the denominators of $\tilde b_1$ and $\tilde b_2$. The resulting
metric function is then given by
\begin{eqnarray}
\label{eq:RSlimit}
e^{4A}&=&4m\left(R_5\over20m^2\right)^{5\over6}(\beta- |y|)\nonumber\\
&=&{4\over L}(\beta-|y|)\ ,
\end{eqnarray}
where $L=m^{-1}(20m^2/R_5)^{5/6}$ and the $y$ coordinate remains
restricted to a compact range,
$|y|<\beta$.  This corresponds to the line element
\begin{equation}
ds^2={2\over\sqrt{L}}(\beta-|y|)^{1\over2}\eta_{\mu\nu}dx^\mu dx^\nu
+ {L^2\over16}{dy^2\over(\beta-|y|)^2}.
\end{equation}
The apparent infinite range of the fifth dimension%
\footnote{As always in anti de Sitter space, proper distance to the
Cauchy horizon is infinite, but the affine parameter along a null
geodesic to the horizon is finite.}
is obtained by making a change of variables
\begin{eqnarray}
\beta-|y|&=&\beta e^{-4|\tilde y|/L},\nonumber\\
x^\mu&=&\left(L\over4\beta\right)^{1\over4}\tilde x^\mu,
\end{eqnarray}
resulting in the five-dimensional metric
\begin{equation}
ds^2 = e^{-2|\tilde y|/L}\eta_{\mu\nu}d\tilde x^\mu d\tilde x^\nu+d\tilde y^2,
\end{equation}
which is literally the Randall-Sundrum solution \cite{RS1,RS2}.  This sign
of the kink ($k<0$) thus corresponds to a binding of gravity to the 3-brane
at $y=0$, with a metric corresponding to segments of pure anti de Sitter
space everywhere off this brane surface.

In taking the above Randall-Sundrum limit $k\rightarrow 0_-$,
$\varphi_0\rightarrow\varphi_*$, the ten-dimensional coordinate $\rho$ is
restricted to a progressively limited range near $\tilde k$, or, equivalently,
$r$ is progressively restricted to a range near $r=0$. Thus, from a $D=10$
perspective, the `infinite' Randall-Sundrum scenario \cite{RS2} corresponds to
shrinking the outer (RS) brane source tightly around the inner horizon brane.
Clearly, what is infinite and what is infinitesimal in this subject is
frame-dependent.

It is instructive to see in addition the scaling of the `brane sources',
(\ref{eq:bsource}), in the Randall-Sundrum limit.  Taking $k\to0_-$, we find
\begin{eqnarray}
T_{MN}^{\rm brane}&=&-{24\over L^2}\delta(y/\beta)\delta_M^\mu
\delta_N^{\nu\vphantom{\mu}} g_{\mu\nu}\nonumber\\
&=&2V_* \delta(y/\beta)\delta_M^\mu \delta_N^{\nu\vphantom{\mu}}
g_{\mu\nu},
\end{eqnarray}
while $Q^{\rm brane}=0$.  This vanishing of the scalar charge is in fact
forced on us since $\varphi$ decouples from the solution in this limit.
This brings up a key observation that it is not so much the breathing
mode $\varphi$ that supports the brane, but rather $H_{[5]}$ flux
corresponding to D3 charge.  In addition, it is also the behavior of
$H_{[5]}$ flux that saves the BPS condition with $Q^{\rm brane}=0$; the
variation $\delta\lambda$ becomes trivial (as it must for a decoupling
scalar), while the gravitino transformation becomes that of pure AdS but with a
sign flip $W_*\to -W_*$ at $y=0$ (corresponding to a Freund-Rubin
compactification with opposite $S^5$ orientations).  This preservation of
supersymmetry further supports the D3-brane origin of the Randall-Sundrum
brane-world, {\it via} the double 3-brane configuration that we have presented.

The above successful reproduction of the Randall-Sundrum scenario
with a `kink-down' (\ie positive tension) domain wall embedded into
$D=5$ anti de Sitter space depends crucially upon use of the breathing
mode $\varphi$, which we have shown to transform in  a necessary
$E_0>4$ anti de Sitter representation. Noted as a possibility for a
Randall-Sundrum scenario in \cite{Cvetic:1999fe,deAlwis:2000pr},
this mode escapes the analysis of \cite{Kallosh:2000tj,BeCvII} because it
belongs to a massive spin-two multiplet, and thus does not belong to
an intrinsically $D=5$ supergravity theory. This is because the full
multiplet of the breathing mode's superpartners cannot be retained in a
`consistent' Kaluza-Klein reduction, since it involves a massive spin
two mode, which never can be kept in a consistent reduction on spheres
\cite{dps}. With respect to the $D=5$, $N=8$ supersymmetry, the breathing
mode belongs to a multiplet containing 20 copies of the following 
sets of fields: 1 spin 2, 4 spin 3/2, 26 spin 1, 20 spin 1/2, 15 spin 0. With respect 
to a $D=5$, $N=2$ decomposition, it belongs to a long massive vector 
supermultiplet\footnote{The $N=8 \rightarrow N=2$ decomposition of 
this bulk multiplet in terms of superconformal $N=4$, $D=4$ boundary 
supermultiplets was given in \cite{Ferrara}.} which is another way of
explaining why it escaped the analysis of \cite{Kallosh:2000tj,BeCvII}.
Since the breathing mode is an $SO(6)$ singlet, only the inclusion of the
breathing mode's non-singlet superpartners leads to difficulties with
Kaluza-Klein consistency; truncation to the purely bosonic theory
involving just $D=5$ gravity and the breathing mode is fully consistent.

\section{Mode locking and spontaneous reduction to an orbifold}

The $Z_2$ symmetric scenario presented above, with two branes of opposite tension
and opposite magnetic charge, corresponding to (\ref{eq:ksol}),
is clearly similar to the brane constructions of
Ho\v{r}ava-Witten orbifolds in M--theory given in \cite{losw1,losw2}. The
analogous type IIB situation has the great advantage that one can work
out explicitly many features of the dynamics, whereas the analogous
discussions in M--theory reduced on Calabi-Yau 3-folds must necessarily
remain rather implicit. Here, we wish to explore further the properties
of this $Z_2$ symmetric solution, and see to which extent it naturally
corresponds to an orbifold compactification.

The orbifold compactification may be viewed as a compactification on
a circle with an additional projection of all the fluctuations into $Z_2$ even
states only. In a Kaluza-Klein spirit, however, one can investigate the
possibility of removing the enforced $Z_2$ projection, in order to see what the
theory does purely of its own accord when compactified about the double 3-brane
background. Thus, we start without making any $Z_2$ projections, but still shall
take the $y$ direction to be a circle.  As explained above, from a
ten-dimensional point of view, the D3-branes now have no noncompact
transverse directions.  Thus there is an added
cohomology constraint, which demands that there cannot be any non-zero net
magnetic charge in the compact transverse space.  Unlike general warped
compactifications, which allow for additional fields and non-trivial
topology, we shall maintain our focus on the round $S^5$ and the breathing
mode of the compactification.  Then, the simplest allowed configuration
on the circle is to have a simple pair of 3-branes with opposite magnetic
charges. Placing the branes at opposite points on the circle gives rise to a
$Z_2$ symmetric configuration.  However, without imposing the $Z_2$ orbifold
symmetry, it would appear that the branes are free to move independently.
But we shall now demonstrate that this is not the case; instead, there is
a mode-locking phenomenon that links the fluctuations of the two 3-branes
into a $Z_2$-invariant combination.

Consider the $y$ coordinate to be periodic with length $2\ell$, making the
identification at $y=\rho_1\leftrightarrow -\rho_2$. For bosonic fields on
this circle, one must impose continuity conditions at both the locations of the
3-branes. Demanding continuity of the scalar field $\varphi$ and the metric
component $e^{2A}$ at $y=0$ and also at $y=\rho_1\leftrightarrow y=-\rho_2$,
one has four continuity conditions to satisfy. In this discussion we shall take
the overall periodicity length $2\ell$ to be fixed, so $\rho_1+\rho_2=2\ell$.
>From continuity of the scalar field $\varphi$, one simply obtains at $y=0$
that the value $\varphi_0$ must be a common limit of $\varphi$ as one approaches
the $y=0$ RS brane either from the left or from the right. Continuity at
$y=\rho_1\leftrightarrow -\rho_2$ implies continuity of the harmonic function
$H$, so one obtains $|k_1|\rho_2=|k_2|\rho_2$, or, using $\rho_1+\rho_2=2\ell$,
that $\left| k_1\over k_2\right|={2\ell\over \rho_1}-1$. Imposing as well
the periodicity conditions on the metric function $e^{2A}$ at $y=0$ and
$y=\rho_1\leftrightarrow -\rho_2$, one obtains the continuity conditions
\be
\label{eq:contconds}
\left|k_2\over k_1\right|
={|m_2|-\sqrt{R_{5(2)}\over 20}e^{{-3\over\sqrt{15}}\varphi_0}
\over |m_1|-\sqrt{{R_{5(1)}\over 20}}e^{{-3\over\sqrt{15}}\varphi_0}}
={|m_2|-\sqrt{R_{5(2)}\over 20}(e^{{-3\over\sqrt{15}}\varphi_0}
+|k_2|\rho_2)
\over |m_1|-\sqrt{{R_{5(1)}\over 20}}(e^{{-3\over\sqrt{15}}\varphi_0}
+|k_1|\rho_1)}\ .
\ee
These conditions are solved by matching relations for $m$ and $R_5$ between
the two regions:
\be
\label{eq:matchingrels}
m_2=({2\ell\over\rho_1}-1)^{-1}m_1\ ,\qquad
\sqrt{R_{5(2)}}=({2\ell\over\rho_1}-1)^{-1}\sqrt{R_{5(1)}}\ .
\ee

Accordingly, if one now makes a standard soliton-physics ansatz
by letting the $Z_2$-odd modulus $\rho_1$ become dependent upon the $D=4$
coordinates $x^\mu$, then upon substitution back into the field equations,
one obtains the effective equation for $\rho_1(x^\mu)$. Because the
oscillations of this coordinate are linked by (\ref{eq:matchingrels})
to the Kaluza-Klein ansatz parameters $m$ and $R_5$, however, this
specific modulus has special restrictions on its oscillations. Both $m$
and $R_5$ are curvature components, and are thus subject to Bianchi
identities. To see this for $m$, consider the Kaluza-Klein ansatz
(\ref{eq:kka}), together with the Bianchi identity
\be
\label{eq:H5bianchi}
dH_{[5]}+\ft12\epsilon_{ij}F^i_{[3]}F^j_{[3]}\equiv 0\ .
\ee
Letting $m\rightarrow m(x)$ and substituting the original ansatz
(\ref{eq:kka}), one obtains directly a suppression of $m$ fluctuations,
$\partial_\mu m(x)=0$. For this reason, parameters entering into
generalized Kaluza-Klein ans\"atze like (\ref{eq:kka}) have been sometimes
been called ``non-zero modes'' \cite{losw2}. In order to see the dynamics
of such modes in more detail, one should restore the massive Kaluza-Klein
modes that are normally set to zero in a compactification. In the case
of $m$, this means replacing the ansatz (\ref{eq:kka}) by
\be
\label{eq:kkamassive}
H_{[5]}=4m(x)e^{8\alpha\varphi}\epsilon_{[5]}
+4m(x)\epsilon_{[5]}(S^5)+h_{[5]}\ ,
\ee
where $h_{[5]}$ represents the fluctuating massive Kaluza-Klein
modes. Re-performing the analysis of the Bianchi identity
(\ref{eq:H5bianchi}) for this generalized ansatz, one now shows
that a non-vanishing $\partial_\mu m$ must be proportional to
$\epsilon^{z_1z_2z_3z_4z_5}\partial_{[z_1}h_{|\mu|z_2z_3z_4z_5]}$,
where $h_{\mu z_2z_3z_4z_5}$ is a Kaluza-Klein massive mode, with mass
determined as usual by the inverse radius of the $S^5$ internal sphere, \ie
corresponding to the length scale of the $D=5$ anti de Sitter space. Thus,
$m(x)$, and hence $\rho_1(x)$ are in fact Kaluza-Klein massive modes,
and become `frozen out' at  energies lower than the AdS scale. Similar
considerations apply to the non-zero mode $R_5$, which is the Ricci scalar
of the internal $S^5$ sphere, upon use of the gravitational curvature
Bianchi identity. Specifically, in the simple case with Kaluza-Klein
massive modes set to zero, if one sets to zero the $D=5$ Bianchi
identity $\nabla^M(R_{MN}-\ft12 g_{MN}R)=0$ and uses the dimensionally
reduced field equations, one finds, for $R_5\rightarrow R_5(x^\mu)$, the
constraint $\partial_\nu R_5\exp(\ft12\sqrt{\ft53}\varphi)-m\partial_\nu
m=0$, thus locking out the low energy $R_5(x^\mu)$ fluctuations as well.

Given that the $Z_2$ odd modes are linked {\it via} Bianchi identities
to massive Kaluza-Klein modes, one expects the theory to settle down
into a low-energy effective theory that is $Z_2$ symmetric. Strictly
speaking, all that has been demonstrated above so far is that the $D=4$
derivatives $\partial_\mu m$, $\partial_\mu R_5$ are locked out at low
energies. In order to show that the theory settles down into a $Z_2$
symmetric lowest energy configuration, one would need either to analyze
in detail the energy functional for the compactified theory, or to
study in more detail the equations of motion of the massive modes. It
is likely that the analysis of $Z_2$ odd modes can only be done fully
consistently if one keeps the entire Kaluza-Klein towers of massive states.

However, one can get an idea of the situation that is obtained with
non-$Z_2$-symmetric configurations if one considers in a little more detail the
question of supersymmetry preservation in a patched background with the matching
conditions (\ref{eq:contconds}, \ref{eq:matchingrels}). Locally, in a patch,
there is no difficulty in finding a Killing spinor. However, once one declares
that the overall compact part of the spacetime is
$S^5\times S^1$, one is required to impose continuity and periodicity
conditions both for bosons and for fermions.

In the $Z_2$ symmetric configuration of the two 3-branes, we have
already demonstrated while discussing the unbroken supersymmetry
transformation of Section \ref{sec:breathingwalls},
\be
\label{eq:goodsusy}
\delta\lambda=\half
e^{-B}(\gamma^{\overline{y}}\varphi'-\varphi')\epsilon\ ,
\ee
that there is a consistently defined and continuous unbroken supersymmetry
transformation with a $Z_2$ even global Killing spinor $\epsilon =
e^{A/2}(1+\gamma^{\overline{y}})\epsilon_0$. Now consider the form
of the {\it broken} supersymmetry transformations in the double
3-brane background. As one can see from the supersymmetry algebra
the anticommutator $\{Q_{\rm broken},Q_{\rm preserved}\}$ involves a
translation in the fifth coordinate $y$, which is clearly $Z_2$ odd. Indeed, the
broken supersymmetry parameters will have $Z_2$ odd projection conditions. This
$Z_2$ odd character is canceled, however, in expressions for Goldstone spinor
zero modes like (\ref{eq:goodsusy}), by the $Z_2$ odd character of $\varphi'$.
Combining the Goldstino expression for $y>0$ with the $Z_2$ map for $y<0$
amounts to inserting an absolute value sign around $\varphi'$ in
(\ref{eq:goodsusy}), taking a broken supersymmetry parameter for $\epsilon$.
Thus, overall, the Goldstino zero mode is $Z_2$ even, as it must be in a
consistent truncation.%
\footnote{The `kink' in the Goldstino expression resulting from
(\ref{eq:goodsusy}) with the replacement $\varphi'\rightarrow |\varphi'|$
corresponds to the sign flip of the superpotential $W$. That $W$ flips
without necessarily passing through zero is what allows the Goldstino mode to
be normalizable in the present case, thus circumventing the normalizability
problems for Goldstinos described in Ref.\ \cite{gibbonslambert}.}
Overall, the zero modes of the double 3-brane geometry
form a single $D=4$, $N=4$ super Maxwell multiplet.

Now consider what happens if one tries to expand around a
non-$Z_2$-symmetric configuration of 3-branes.  For the Killing spinor itself, 
one may observe that $\epsilon =
e^{A/2}(1+\gamma^{\overline{y}})\epsilon_0$ is in fact still continuous and
well-behaved in the non-symmetric case, since the metric function
$e^{2A}$ is by construction matched at the branes.  However the situation is
different for the candidate Goldstinos. For a non-$Z_2$-symmetric configuration
the derivative $\varphi'$ differs by more than a sign as one crosses a 3-brane:
in this case one has $|k_2|\ne|k_1|$, so there is a non-unimodular factor
present as well. This prevents one from having continuity both of the unbroken
supersymmetry parameter and of the Goldstinos. Thus, although things
look locally like one has a BPS configuration with unbroken supersymmetry for a
non-$Z_2$-symmetric configuration, analysis of the putative zero-mode
supermultiplets finds them to be inconsistent with the available matching
conditions. So, one is lead to conclude that only the $Z_2$ symmetric
configuration has a proper unbroken supersymmetry and zero-mode multiplets
transforming correctly with respect to it.

The configuration with globally unbroken supersymmetry should be the proper
`vacuum' in this double 3-brane sector of type IIB theory compactified
on $S^5$. A fuller analysis of this spontaneous reduction to a $Z_2$
invariant effective theory on the basis of energy functionals and
the equations of motion for the Kaluza-Klein massive modes would
be desirable. But it is already clear that this double 3-brane model
displays a remarkable spontaneous appearance of an orbifold structure. This
happens not by insistent projection into a $Z_2$ invariant sector of the theory,
but naturally by virtue of the Kaluza-Klein dynamics of the theory.

Our discussion has indicated that the original Randall-Sundrum model
\cite{RS1} arises naturally when the fifth dimension $y$ direction is
taken to be compact, and one may view the model as a system of two D3-branes
transverse to the internal $S_1\times S^5$.  From the $D=5$ point of view,
there are two branes, one with positive and one with negative tension,
constrained by Kaluza-Klein dynamics to live at diametrically opposed
points on the circle.  While the presence of a negative tension brane
might appear troublesome, we have shown that it does not contribute the
na\"\i{v}ely anticipated negative energy modes; these are non-zero modes
and mix with higher Kaluza-Klein massive modes. The negative tension
3-brane has the effect of protecting the spacetime from curvature
singularities in the geometry that might reside behind the Cauchy
horizon. Of the {\it a priori} two independent types of motion of the 3-branes
along the $S^1$ direction, only the $Z_2$ even modes, corresponding to
an overall `rotation' of both branes along the circle, localized in the
$D=4$ coordinates $x^\mu$, correspond to genuine zero modes.

\section{Conclusions}

We have found that an appropriately constructed D3-brane configuration
provides a supersymmetric and dynamically stable Randall-Sundrum scenario. This
is achieved in a solution to the $D=10$ type IIB supergravity equations which
can be given a $D=5$ interpretation, but is not fully a $D=5$ solution, for it
employs an intrinsically massive Kaluza-Klein mode, the $S^5$ breathing 
mode. This mode has AdS energy $E_0=8$, satisfying the bound 
$E_{0}>4$ that is required for an asymptotic approach to AdS space 
from a downwards-facing warp-factor kink in a Randall-Sundrum scenario.
There is also a $Z_2$ flip in the sign of the Freund-Rubin parameter 
$m$. This is natural enough in a $D=10$ context where $m$ is a field-strength
value, but
it is less natural from a $D=5$ viewpoint, where $m$ normally would appear as a
parameter. We have found, moreover, that although one can decide to exclude the
$Z_2$ odd modes when expanding the theory around the presented $Z_2$ invariant background,
and thus reproduce an $S^1/Z_2$ orbifold reduction, it is not
actually necessary to make this projection by hand. Bianchi identities
for the curvature values entering in the solution relate the
$Z_2$ odd modes to Kaluza-Klein massive states of the theory, and so
they decouple naturally at low energy. Although charge conservation on the
circle
requires branes to come in oppositely charged pairs, we have seen that one can
recover a single brane Randall-Sundrum model by pushing the second brane off to
the Cauchy horizon (\ie by taking $\varphi_1=\varphi_*$ for the second brane).
{}From the $D=10$ point of view, however, this corresponds to shrinking an outer
RS shell of D3 brane tightly around an inner `horizon' D3 brane of opposite
charge and tension. Clearly, an important problem is whether this
geometry can be realized in a string theory context.

\section*{Acknowledgments}
We would like to acknowledge stimulating discussions with Shanta
de~Alwis, Mirjam Cvetic, Marc Henneaux, Renata Kallosh, Neil Lambert,
Finn Larsen, Ruben Minasian, Herve Partouche, Chris Pope, Lisa 
Randall, Hisham Sati and Herman Verlinde. 

\section*{Note added}
As this paper was in the final stages of preparation, a very interesting paper
appeared \cite{bkvp} that sheds light on the relationship between constructions
such as those of Refs.\ \cite{losw1,losw2} or the present paper and the
supersymmetry scheme of Ref.\ \cite{abn}, which was otherwise puzzling.
Ref.\ \cite{bkvp} discusses supersymmetry in orbifolds, and in particular the
$D=5$ case of interest here. In order to obtain a continuous Killing spinor
at orbifold singularities (necessary for the Killing equation to be realized
everywhere, including at the singular points), Ref.\ \cite{bkvp} introduces a
5-form `theory of nothing' field strength, which has just a constant as a
solution, but allows for this variable to be only piecewise constant. This
allows for a $Z_2$ sign flip in the prepotential that is critical for having a
preserved supersymmetry allowing coupling to supermatter. This sign flip was
not made in the discussion of Ref.\ \cite{abn}, leading to problems with matter
coupling.  This difficulty of \cite{abn}, and the resolution of \cite{bkvp}
was also investigated in \cite{Falkowski:2000er,Alonso-Alberca:2000ne} and
independently worked out by \cite{partouche}.
We anticipate that a fuller Kaluza-Klein treatment of the type IIB
theory, including all fermions and making a careful reduction of the type IIB
supersymmetry transformations, will show that the $D=5$ supersymmetry
realization adopted in Ref.\ \cite{bkvp} can also be viewed as the natural
dimensional reduction of the type IIB theory using a $Z_2$ symmetrized ansatz
of the type employed in Refs.\ \cite{losw1,losw2} and the present paper.
In particular, we expect that the 5-form `theory of nothing' field introduced
in Refs.\ \cite{Alonso-Alberca:2000ne,bkvp} can be identified with the $D=5$
residue of the type IIB self-dual 5-form field strength.


\begin{thebibliography}{99}


\bibitem{Bremer}
M.~S.~Bremer, M.~J.~Duff, H.~L\"u, C.~N.~Pope and K.~S.~Stelle,
{\sl Instanton cosmology and domain walls from M-theory and string theory},
Nucl.\ Phys.\ {\bf B543}, 321 (1999) [hep-th/9807051].

\bibitem{RS1}
L.~Randall and R.~Sundrum,
{\sl A large mass hierarchy from a small extra dimension},
Phys.\ Rev.\ Lett.\ {\bf 83}, 3370 (1999) [hep-ph/9905221].

\bibitem{RS2}
L.~Randall and R.~Sundrum,
{\sl An alternative to compactification},
Phys.\ Rev.\ Lett.\ {\bf 83}, 4690 (1999) [hep-th/9906064].

\bibitem{Akama}
K.~Akama,
{\sl Pregeometry},
Lect. Notes Phys. {\bf 176}, 267 (1982) [hep-th/0001113].

\bibitem{Rubakov:1983bb}
V.~A.~Rubakov and M.~E.~Shaposhnikov,
{\sl Do We Live Inside A Domain Wall?},
Phys.\ Lett.\ {\bf B125}, 136 (1983).

\bibitem{Rubakov:1983bz}
V.~A.~Rubakov and M.~E.~Shaposhnikov,
{\sl Extra Space-Time Dimensions: Towards A Solution To The Cosmological
Constant Problem},
Phys.\ Lett.\ {\bf B125}, 139 (1983).

\bibitem{Gibbons:1988ps}
G.~W.~Gibbons and K.~Maeda,
{\sl Black Holes And Membranes In Higher Dimensional Theories With Dilaton
Fields},
Nucl.\ Phys.\ {\bf B298}, 741 (1988).

\bibitem{Horowitz:1991cd}
G.~T.~Horowitz and A.~Strominger,
{\sl Black strings and P-branes},
Nucl.\ Phys.\ {\bf B360}, 197 (1991).

\bibitem{Duff:1991pe}
M.~J.~Duff and J.~X.~Lu,
{\sl The Selfdual type IIB superthreebrane},
Phys.\ Lett.\ {\bf B273}, 409 (1991).

\bibitem{Horava:1996qa}
P.~Ho\v{r}ava and E.~Witten,
{\sl Heterotic and type I string dynamics from eleven dimensions},
Nucl.\ Phys.\ {\bf B460}, 506 (1996) [hep-th/9510209].

\bibitem{Horava:1996ma}
P.~Hora\v{v}a and E.~Witten,
{\sl Eleven-Dimensional Supergravity on a Manifold with Boundary},
Nucl.\ Phys.\ {\bf B475}, 94 (1996) [hep-th/9603142].

\bibitem{losw1}
A. Lukas, B. Ovrut, K. S. Stelle and D. Waldram,
{\sl The Universe as a Domain Wall},
Phys.\ Rev.\ {\bf D59}, 086001 (1999) [hep-th/9803235].

\bibitem{losw2}
A. Lukas, B. Ovrut, K. S. Stelle and D. Waldram,
{\sl Heterotic M--theory in Five Dimensions},
Nucl.\ Phys.\ {\bf B552}, 246 (1999) [hep-th/9806051].

\bibitem{ls}
A. Lukas and K. S. Stelle,
{\sl Heterotic anomaly cancellation in five dimensions},
JHEP {\bf 0001}, 010 (2000) [hep-th/9911156].

\bibitem{Youm:2000bz}
D.~Youm,
{\sl Solitons in brane worlds},
Nucl.\ Phys.\ {\bf B576}, 106 (2000) [hep-th/9911218].

\bibitem{Youm:2000zp}
D.~Youm,
{\sl Probing solitons in brane worlds},
Nucl.\ Phys.\ {\bf B576}, 123 (2000) [hep-th/9912175].

\bibitem{Cvetic:1999fe}
M.~Cveti\v{c}, H.~L\"u and C.~N.~Pope,
{\sl Domain walls and massive gauged supergravity potentials},
hep-th/0001002.

\bibitem{Cvetic:2000dz}
M.~Cveti\v{c}, H.~L\"u and C.~N.~Pope,
{\sl Localised gravity in the singular domain wall background?},
hep-th/0002054.

\bibitem{deAlwis:2000qc}
S.~P.~de Alwis,
{\sl Brane world scenarios and the cosmological constant},
hep-th/0002174.

\bibitem{deAlwis:2000pr}
S.~P.~de Alwis, A.~T.~Flournoy and N.~Irges,
{\sl Brane worlds, the cosmological constant and string theory},
hep-th/0004125.

\bibitem{Duffliu}
M. J. Duff and James T. Liu, 
{\sl Complementarity of the Maldacena and Randall-Sundrum pictures},
Phys. Rev. Lett. {\bf 85}, 2052 (2000) [hep-th 0003237].

\bibitem{Kallosh:2000tj}
R.~Kallosh and A.~Linde,
{\sl Supersymmetry and the brane world},
JHEP {\bf 0002}, 005 (2000) [hep-th/0001071].

\bibitem{BeCvII}
K.~Behrndt and M.~Cveti\v{c},
{\sl Anti-de Sitter vacua of gauged supergravities with 8 supercharges},
Phys.\ Rev.\ {\bf D61}, 101901 (2000) [hep-th/0001159].

\bibitem{Ceresole:2000jd}
A.~Ceresole and G.~Dall'Agata,
{\sl General matter coupled ${\cal N}=2$, $D=5$ gauged supergravity},
hep-th/0004111.

\bibitem{maldacena}
J.~Maldacena and C.~Nu\~nez,
{\sl Supergravity description of field theories on curved manifolds and a
no-go theorem},
hep-th/0007018.

\bibitem{Wijnholt:1999vk}
M.~Wijnholt and S.~Zhukov,
{\sl On the uniqueness of black hole attractors},
hep-th/9912002.

\bibitem{Kraus:1999it}
P.~Kraus,
{\sl Dynamics of anti-de Sitter domain walls},
JHEP {\bf 9912}, 011 (1999) [hep-th/9910149].

\bibitem{Chan:2000ms}
C.~S.~Chan, P.~L.~Paul and H.~Verlinde,
{\sl A note on warped string compactification},
Nucl.\ Phys.\ {\bf B581}, 156 (2000) [hep-th/0003236].

\bibitem{Mayr:2000zd}
P.~Mayr,
{\sl Stringy world branes and exponential hierarchies},
hep-th/0006204.

\bibitem{Skenderis:1999mm}
K.~Skenderis and P.~K.~Townsend,
{\sl Gravitational stability and renormalization-group flow},
Phys.\ Lett.\ {\bf B468}, 46 (1999) [hep-th/9909070].

\bibitem{Freedman:1999gp}
D.~Z.~Freedman, S.~S.~Gubser, K.~Pilch and N.~P.~Warner,
{\sl Renormalization group flows from holography---supersymmetry and a
c-theorem},
hep-th/9904017.

\bibitem{Behrndt:2000kz}
K.~Behrndt and M.~Cveti\v{c},
{\sl Supersymmetric domain wall world from $D=5$ simple gauged supergravity},
Phys.\ Lett.\ {\bf B475}, 253 (2000) [hep-th/9909058].

\bibitem{DeWolfe:1999cp}
O.~DeWolfe, D.~Z.~Freedman, S.~S.~Gubser and A.~Karch,
{\sl Modeling the fifth dimension with scalars and gravity},
Phys.\ Rev.\ {\bf D62}, 046008 (2000) [hep-th/9909134].

\bibitem{Flato:1984te}
M.~Flato and C.~Fronsdal,
{\sl Representations Of Conformal Supersymmetry},
Lett.\ Math.\ Phys.\  {\bf 8}, 159 (1984).

\bibitem{Dobrev:1985qv}
V.~K.~Dobrev and V.~B.~Petkova,
{\sl All Positive Energy Unitary Irreducible Representations Of Extended
Conformal Supersymmetry},
Phys.\ Lett.\  {\bf B162}, 127 (1985).

\bibitem{Bars:1983ep}
I.~Bars and M.~Gunaydin,
{\sl Unitary Representations Of Noncompact Supergroups},
Commun.\ Math.\ Phys.\  {\bf 91}, 31 (1983).

\bibitem{Gunaydin:1988hb}
M.~Gunaydin,
{\sl Unitary Highest Weight Representations Of Noncompact Supergroups},
J.\ Math.\ Phys.\  {\bf 29}, 1275 (1988).

\bibitem{Gunaydin:1999jc}
M.~Gunaydin, D.~Minic and M.~Zagermann,
{\sl Novel supermultiplets of $SU(2,2|4)$ and the AdS$_5$/CFT$_4$ duality},
Nucl.\ Phys.\  {\bf B544}, 737 (1999)
[hep-th/9810226].

\bibitem{Gunaydin:1998sw}
M.~Gunaydin, D.~Minic and M.~Zagermann,
{\sl 4D doubleton conformal theories, $CPT$ and IIB strings on AdS$_5
\times S^5$},
Nucl.\ Phys.\  {\bf B534}, 96 (1998) [hep-th/9806042].

\bibitem{Ferrara}
S.~Ferrara. M.~A.~Lledo and A.~Zaffaroni,
{\sl Born-Infeld corrections to D3 brane action in AdS$_5 \times\rm S_5$ and 
$N=4$, $D=4$ primary superfields},
Phys.\ Rev.\ {\bf D58}, 105029 (1998) [hep-th/9805082].

\bibitem{Breitenlohner:1982jf}
P.~Breitenlohner and D.~Z.~Freedman,
{\sl Stability In Gauged Extended Supergravity},
Annals Phys.\ {\bf 144}, 249 (1982).

\bibitem{deBoer:1999xf}
J.~de Boer, E.~Verlinde and H.~Verlinde,
{\sl On the holographic renormalization group},
JHEP {\bf 0008}, 003 (2000) [hep-th/9912012].

\bibitem{Gubser:1998bc}
S.~S.~Gubser, I.~R.~Klebanov and A.~M.~Polyakov,
{\sl Gauge theory correlators from non-critical string theory},
Phys.\ Lett.\  {\bf B428}, 105 (1998) [hep-th/9802109].

\bibitem{Witten:1998qj}
E.~Witten,
{\sl Anti-de Sitter space and holography},
Adv.\ Theor.\ Math.\ Phys.\  {\bf 2}, 253 (1998) [hep-th/9802150].

\bibitem{kksugra}
M.~J.~Duff, B.~E.~W.~Nilsson and C.~N.~Pope,
{\sl Kaluza-Klein Supergravity},
Phys. Rep. {\bf 130}, 1 (1986).

\bibitem{Cvetic:2000dm}
M.~Cveti\v{c}, H.~L\"u and C.~N.~Pope,
{\sl Consistent Kaluza-Klein sphere reductions},
Phys.\ Rev.\ {\bf D62}, 064028 (2000) [hep-th/0003286].

\bibitem{Duff:1995an}
M.~J.~Duff, R.~R.~Khuri and J.~X.~Lu,
{\sl String solitons},
Phys.\ Rept.\ {\bf 259}, 213 (1995) [hep-th/9412184].

\bibitem{Duff:1998me}
M.~J.~Duff, J.~T.~Liu, H.~L\"u and C.~N.~Pope,
{\sl Gauge dyonic strings and their global limit},
Nucl.\ Phys.\ {\bf B529}, 137 (1998) [hep-th/9711089].

\bibitem{wip} M. Cveti\v{c}, M. J. Duff, J. T. Liu, H. L\"u, C. N. Pope
and K. S. Stelle, work in progress.

\bibitem{dps} M.~J.~Duff, C.~N.~Pope, and K.~S.~Stelle,
{\sl Consistent interacting massive spin-2 requires an infinity of states},
Phys.\ Lett.\ {\bf B223}, 386 (1989).

\bibitem{gibbonslambert} G.~W.~Gibbons and N.~D.~Lambert,
{\sl Domain walls and solitons in odd dimensions},
Phys.\ Lett.\ {\bf B488}, 90 (2000) [hep-th/0003197].

\bibitem{bkvp} E.~Bergshoeff, R.~Kallosh and A.~Van Proeyen,
{\sl Supersymmetry in singular spaces},
hep-th/0007044.

\bibitem{abn} R.~Altendorfer, J.~Bagger and D.~Nemeschansky,
{\sl Supersymmetric Randall-Sundrum scenario},
hep-th/0003117.

\bibitem{Falkowski:2000er}
A.~Falkowski, Z.~Lalak and S.~Pokorski,
{\sl Supersymmetrizing branes with bulk in five-dimensional supergravity},
hep-th/0004093.

\bibitem{Alonso-Alberca:2000ne}
N.~Alonso-Alberca, P.~Meessen and T.~Ortin,
{\sl Supersymmetric brane-worlds},
Phys.\ Lett.\ {\bf B482}, 400 (2000) [hep-th/0003248].

\bibitem{partouche} J. T. Liu, R. Minasian and H. Partouche, unpublished.

\end{thebibliography}
\end{document}